\documentclass[aps,showpacs,twocolumn]{revtex4}
\usepackage{graphicx}

\begin{document}


\title{Electromagnetic signatures of thin accretion disks in
wormhole geometries}

\author{Tiberiu Harko}
\email{harko@hkucc.hku.hk} \affiliation{Department of Physics and
Center for Theoretical and Computational Physics, The University
of Hong Kong, Pok Fu Lam Road, Hong Kong}

\author{Zolt\'{a}n Kov\'{a}cs}
\email{zkovacs@mpifr-bonn.mpg.de}

\affiliation{Max-Planck-Institute f\"{u}r Radioastronomie, Auf dem
H\"{u}gel 69, 53121 Bonn, Germany}
\affiliation{Department of Experimental Physics, University of
Szeged, D\'{o}m T\'{e}r 9, Szeged 6720, Hungary}

\author{Francisco S. N. Lobo}
\email{francisco.lobo@port.ac.uk} \affiliation{Institute of
Cosmology \& Gravitation,
             University of Portsmouth, Portsmouth PO1 2EG, UK}
\affiliation{Centro de Astronomia e Astrof\'{\i}sica da
Universidade de Lisboa, Campo Grande, Ed. C8 1749-016 Lisboa,
Portugal}

\date{\today}

\begin{abstract}

In this paper, we study the physical properties and
characteristics of matter forming thin accretion disks in static
and spherically symmetric wormhole spacetimes. In particular, the
time averaged energy flux, the disk temperature and the emission
spectra of the accretion disks are obtained for these exotic
geometries, and are compared with the Schwarzschild solution. It
is shown that more energy is emitted from the disk in a wormhole
geometry than in the case of the Schwarzschild potential and the
conversion efficiency of the accreted mass into radiation is more
than a factor of two higher for the wormholes than for static
black holes. These effects in the disk radiation are confirmed in
the radial profiles of temperature corresponding to theses flux
distributions, and in the emission spectrum $\omega L(\omega)$ of
the accretion disks. We conclude that specific signatures appear
in the electromagnetic spectrum, thus leading to the possibility
of distinguishing wormhole geometries by using astrophysical
observations of the emission spectra from accretion disks.

\end{abstract}

\pacs{04.50.Kd, 04.70.Bw, 97.10.Gz}

\maketitle

\section{Introduction}

Most of the astrophysical bodies grow substantially in mass via
accretion. Recent observations suggest that around almost all of
the active galactic nuclei (AGN's), or black hole candidates,
there exist gas clouds surrounding the central compact object,
together with an associated accretion disk, on a variety of scales
from a tenth of a parsec to a few hundred parsecs \cite{UrPa95}.
These gas clouds, existing in either a molecular or atomic phase,
are assumed to form a geometrically and optically thick torus (or
warped disk), which absorbs most of the ultraviolet radiation and
soft X-rays. The most powerful evidence for the existence of super
massive black holes comes from the VLBI (very long baseline
interferometry) imaging of molecular $\mathrm{H_{2}O}$ masers in
the active galaxy NGC 4258 \cite {Miyo95}. This imaging, produced
by Doppler shift measurements assuming Keplerian motion of the
masering source, has allowed a quite accurate estimation of the
central mass, which has been found to be a $3.6\times
10^{7}M_{\odot }$ super massive dark object, within $0.13$
parsecs. Hence, important astrophysical information can be
obtained from the observation of the motion of gas streams in the
gravitational field of compact objects.

The mass accretion around rotating black holes was studied in
general relativity for the first time in \cite{NoTh73}.  By using
an equatorial approximation to the stationary and axisymmetric
spacetime of rotating black holes, steady-state thin disk models
were constructed, extending the theory of non-relativistic
accretion \cite{ShSu73}. In these models hydrodynamical
equilibrium is maintained by efficient cooling mechanisms via
radiation transport, and the accreting matter has a Keplerian
rotation. The radiation emitted by the disk surface was also
studied under the assumption that black body radiation would
emerge from the disk in thermodynamical equilibrium. The radiation
properties of thin accretion disks were further analyzed  in
 \cite{PaTh74,Th74}, where the effects of photon
capture by the hole on the spin evolution were presented as well.
In these works the efficiency with which black holes convert rest
mass into outgoing radiation in the accretion process was also
computed. More recently, the emissivity properties of the
accretion disks were investigated for exotic central objects, such
as non-rotating or rotating quark, boson or fermion stars
\cite{Bom,To02,YuNaRe04,Guzman:2005bs}. The radiation power per
unit area, the temperature of the disk and the spectrum of the
emitted radiation were given, and compared with the case of a
Schwarzschild black hole of an equal mass. The physical properties
of matter forming a thin accretion disk in the static and
spherically symmetric spacetime metric of vacuum $f(R)$ modified
gravity models were also analyzed \cite{Pun:2008ae}, and it was
shown that particular signatures can appear in the electromagnetic
spectrum, thus leading to the possibility of directly testing
modified gravity models by using astrophysical observations of the
emission spectra from accretion disks.

It is the purpose of the present paper to study the thin accretion
disk models applied for the study of wormholes, and carry out an
analysis of the properties of the radiation emerging from the
surface of the disk. In classical general relativity, wormholes
are supported by exotic matter, which involves a stress energy
tensor that violates the null energy condition (NEC)
\cite{Morris:1988cz}. Note that the NEC is given by
$T_{\mu\nu}k^\mu k^\nu \geq 0$, where $k^\mu$ is {\it any} null
vector. Several candidates have been proposed in the literature,
amongst which we refer to the first solutions with what we now
call massless phantom scalar fields \cite{Ellis-Bronn} and
(probably) the first examples of multidimensional wormhole
solutions \cite{Clement}; solutions in higher dimensions, for
instance in Einstein-Gauss-Bonnet theory \cite{EGB}; wormholes on
the brane \cite{braneWH}; solutions in Brans-Dicke theory
\cite{Nandi:1997en}; wormhole solutions in semi-classical gravity
(see Ref. \cite{Garattini:2007ff} and references therein); exact
wormhole solutions using a more systematic geometric approach were
found \cite{Boehmer:2007rm}; solutions supported by equations of
state responsible for the cosmic acceleration \cite{phantomWH};
and NEC respecting geometries were further explored in conformal
Weyl gravity \cite{Lobo:2008zu}, etc (see Refs.
\cite{Lemos:2003jb,Lobo:2007zb} for more details and
\cite{Lobo:2007zb} for a recent review). Thus, it should prove
interesting to analyze the properties of an accretion disk around
wormholes comprising of exotic matter. In fact, we show that
specific signatures appear in the electromagnetic spectrum, thus
leading to the possibility of distinguishing wormhole geometries
from the Schwarzschild solution by using astrophysical
observations of the emission spectra from accretion disks.

The present paper is organized as follows.  In Section
\ref{sec:II}, we review the formalism of the thin disk accretion
onto compact objects for static and spherically symmetric
spacetimes. In Section \ref{sec:III}, we analyze the basic
properties of matter forming a thin accretion disk in wormhole
spacetimes. We discuss and conclude our results in Section
\ref{sec:concl}. Throughout this work, we use a system of units so
that $c=G=\hbar =k_{B}=1$, where $k_{B}$ is Boltzmann's constant.

\section{Electromagnetic radiation properties of thin accretion
disks in general relativistic spacetimes}\label{sec:II}

Accretion discs are flattened astronomical objects made of rapidly rotating gas which slowly spirals onto a central gravitating body, with its gravitational energy degraded to heat. A fraction of the heat converts into radiation, which partially escapes, and cools down the accretion disc.
The only information that we have about accretion disk physics comes from this radiation, when it reaches radio, optical and $X$-ray telescopes, allowing astronomers to analyze its electromagnetic spectrum, and its time variability. A thin accretion disk is an accretion disk such that in cylindrical coordinates $(r,\phi ,z)$ most of the matter lies close to the radial plane.
For the thin accretion disk its vertical size (defined along the $z$-axis)
is negligible, as compared to its horizontal extension (defined along the radial direction $r$), i.e, the disk height $H$, equal to the maximum half thickness of the
disk, is always much smaller than the characteristic radius $R$ of
the disk, $H \ll R$.

The thin disk is in hydrodynamical equilibrium, and the pressure gradient and a vertical entropy
gradient in the accreting matter are negligible. The efficient
cooling via the radiation over the disk surface prevents the disk
from cumulating the heat generated by stresses and dynamical
friction. In turn, this equilibrium causes the disk to stabilize
its thin vertical size. The thin disk has an inner edge at the
marginally stable orbit of the compact object potential, and the
accreting plasma has a Keplerian motion in higher orbits.

In steady state accretion disk models, the mass accretion rate
$\dot{M}_{0}$ is assumed to be a constant that does not change
with time. The physical quantities describing the orbiting
plasma are averaged over a characteristic time scale, e.g. $\Delta
t$, over the azimuthal angle $\Delta \phi =2\pi $ for a total
period of the orbits, and over the height $H$ \cite{ShSu73,
NoTh73,PaTh74}.

The particles moving in Keplerian orbits around the compact object
with a rotational velocity $\Omega =d\phi /dt$ have a specific
energy $\widetilde{E} $ and a specific angular momentum
$\widetilde{L\text{,}}$ which, in the steady state thin disk
model, depend only on the radii of the orbits. These particles,
orbiting with the four-velocity $u^{\mu }$, form a disk of an
averaged surface density $\Sigma $, the vertically integrated
average of the rest mass density $\rho _{0}$ of the plasma. The
accreting matter in the disk is modelled by an anisotropic fluid
source, where the density $\rho _{0}$, the energy flow vector
$q^{\mu }$ and the stress tensor $t^{\mu \nu }$ are measured in
the averaged rest-frame (the specific heat was neglected). Then,
the disk structure can be characterized by the surface density of
the disk \cite{NoTh73,PaTh74},
\begin{equation}
\Sigma(r) = \int^H_{-H}\langle\rho_0\rangle dz,
\end{equation}
with averaged rest mass density $\langle\rho_0\rangle$ over
$\Delta t$ and $ 2\pi$ and the torque
\begin{equation}
W_{\phi}{}^{r} =\int^H_{-H}\langle t_{\phi}{}^{r}\rangle dz,
\end{equation}
with the averaged component $\langle t^r_{\phi} \rangle$ over
$\Delta t$ and $2\pi$. The time and orbital average of the energy
flux vector gives the radiation flux ${\mathcal F}(r)$ over the
disk surface as ${\mathcal F}(r)=\langle q^z \rangle$.

The stress-energy tensor is decomposed according to
\begin{equation}
T^{\mu \nu }=\rho_{0}u^{\mu }u^{\nu }+2u^{(\mu }q^{\nu )}+t^{\mu
\nu },
\end{equation}
where $u_{\mu }q^{\mu }=0$, $u_{\mu }t^{\mu \nu }=0$. The
four-vectors of the energy and angular momentum flux are defined
by $-E^{\mu }\equiv T_{\nu }^{\mu }{}(\partial /\partial t)^{\nu
}$ and $J^{\mu }\equiv T_{\nu }^{\mu }{}(\partial /\partial \phi
)^{\nu }$, respectively. The structure equations of the thin disk
can be derived by integrating the conservation laws of the rest
mass, of the energy, and of the angular momentum of the plasma
\cite{NoTh73,PaTh74}. From the equation of the rest mass
conservation, $(\rho _{0}u^{\mu })_{;\mu }=0$, it follows that the
time averaged rate of the accretion of the rest mass is
independent of the disk radius,
\begin{equation}
\dot{M_{0}}\equiv -2\pi r\Sigma u^{r}={\rm constant}.
\label{conslawofM}
\end{equation}

The conservation law $E^{\mu }{}_{\mu }=0$ of the energy has the
integral form
\begin{equation}
\lbrack \dot{M}_{0}\widetilde{E}-2\pi r\Omega W_{\phi }{}^{r}]_{,r}
=4\pi r{\mathcal F}%
\widetilde{E}\;\;,  \label{conslawofE}
\end{equation}
which states that the energy transported by the rest mass flow,
$\dot{M}_{0} \widetilde{E}$, and transported by the dynamical
stresses in the disk, $2\pi r\Omega W_{\phi }{}^{r}$, is in
balance with the energy radiated away from the surface of the
disk, $4\pi r{\mathcal F}\widetilde{E}$. The law of the angular
momentum conservation, $J^{\mu }{}_{\mu }=0$, also states the
balance of these three forms of the angular momentum transport,
\begin{equation}
\lbrack \dot{M}_{0}\widetilde{L}-2\pi rW_{\phi }{}^{r}]_{,r}=4\pi
r{\mathcal F} \widetilde{L}\;\;.  \label{conslawofL}
\end{equation}

By eliminating $W_{\phi }{}^{r}$ from Eqs. (\ref{conslawofE}) and
(\ref {conslawofL}), and applying the universal energy-angular
momentum relation $ dE=\Omega dJ$ for circular geodesic orbits in
the form $\widetilde{E}_{,r}=\Omega \widetilde{L}_{,r}$, the flux
${\mathcal F}$ of the radiant energy over the disk can be
expressed in terms of the specific energy, angular momentum and of
the angular velocity of the black hole \cite{NoTh73,PaTh74},
\begin{equation}
{\mathcal F}(r)=-\frac{\dot{M}_{0}}{4\pi \sqrt{-g}}\frac{\Omega
_{,r}}{(\widetilde{E}-\Omega
\widetilde{L})^{2}}\int_{r_{ms}}^{r}(\widetilde{E}-\Omega
\widetilde{L}) \widetilde{L}_{,r}dr\;\;.  \label{F}
\end{equation}

In the derivation of the above formula the ``no torque'' inner
boundary condition was prescribed, according to which the torque
vanishes at the inner edge of the disk. Thus, we assume that the
accreting matter at the marginally stable orbit $r_{ms}$ falls
freely into the central compact object, and
cannot exert considerable torque on the disk. The latter
assumption is valid only if no strong magnetic fields exist in
the plunging region, where matter falls into the compact object.

Another important characteristics of the mass accretion process is
the efficiency with which the central object converts rest mass
into outgoing radiation. This quantity is defined as the ratio of
the rate of the radiation energy of photons escaping from the
disk surface to infinity, and the rate at which mass-energy is
transported to the central compact general relativistic object,
both measured at infinity \cite{NoTh73,PaTh74}. If all the emitted
photons can escape to infinity, the efficiency is given in terms
of the specific energy measured at the marginally stable orbit
$r_{ms}$,
\begin{equation}
\epsilon =1-\widetilde{E}_{ms}.\label{epsilon}
\end{equation}

For Schwarzschild black holes the efficiency $\epsilon $ is about
$6\%$, whether the photon capture by the black hole is considered,
or not. Ignoring the capture of radiation by the hole, $\epsilon $
is found to be $42\%$ for rapidly rotating black holes, whereas
the efficiency is $40\%$ with photon capture in the Kerr potential
\cite{Th74}.

It is possible to define a temperature $T(r)$ of the disk, by
using the definition of the flux, as ${\mathcal F}(r)=\sigma
T^{4}(r)$, where $\sigma $ is the Stefan-Boltzmann constant.
Considering that the disk emits as a black body, one can use the
dependence of $T$ on ${\mathcal F}$ to calculate the luminosity
$L\left(\omega \right) $ of the disk through the expression for
the black body spectral distribution \cite{To02},
\begin{equation}
L\left(\omega \right) =4\pi d^{2}I\left(\omega
\right)=\frac{4}{\pi }\cos i \omega
^{3}\int_{r_{i}}^{r_{f}}\frac{rdr}{\exp \left(\omega /T\right)
-1},
\end{equation}
where $d$ is the distance to the source, $\omega=2\pi \nu$, where
$\nu $ is the radiation frequency, $I(\omega)$ denotes the Planck
distribution function, $i$ is the inclination of the accretion disk (defined as the angle between the line of sight and the normal to the disk), and $r_{i}$
and $r_{f}$ indicate the position of the inner and outer edge of
the disk, respectively. In the following the inclination angle $i
$ used for the calculation of the spectra is set to $\cos i =1$.

In order to compute the flux integral given by Eq.~(\ref{F}), we
determine the radial dependence of the angular velocity $\Omega$,
of the specific energy $\widetilde{E}$ and of the specific angular
momentum $\widetilde{L}$ of particles moving in circular orbits
around general relativistic compact spheres in a static and
spherically symmetric geometry given by the following line element
\begin{equation}\label{metr1}
ds^2=-e^{2\Phi (r)}dt^2+e^{2\lambda(r)}dr^2
+r^2(d\theta^2+\sin^2\theta\,d\phi^2)\,.
\end{equation}

The geodesic equations for particles orbiting in the equatorial
plane of the compact object can be written as
\begin{eqnarray}
e^{4\Phi }\left(\frac{d{t}}{d\tau}\right)^2 = \widetilde{E}^2,
\quad r^4\left(\frac{d\phi}{d\tau}\right)^2 =
\widetilde{L}^2, \\
e^{4\left(\Phi +\lambda \right)}\left(\frac{d{r}}{d\tau}\right)^2+
V_{eff}(r)=\widetilde E^2,
\end{eqnarray}
where $\tau$ is the affine parameter, and the effective potential
is given by
\begin{equation}
V_{eff}(r)\equiv e^{2\Phi
}\left(1+\frac{\widetilde{L}^2}{r^2}\right)\;. \label{V2}
\end{equation}
From the conditions $V_{eff}(r)=0$ and
$V_{eff,\left.\right.r}(r)=0$ we obtain
\begin{eqnarray}
\widetilde{E}&=& \frac{e^{2\Phi}}{\sqrt{e^{2\Phi}-r^2\Omega^2}}\;,
      \label{E}    \\
\widetilde{L}&=&\frac{r^2\Omega}{\sqrt{e^{2\Phi}-r^2\Omega^2}}\;.
      \label{L}    \\
\Omega&=&\sqrt{\frac{\Phi_{,r}e^{2\Phi }}{r}}\;,
      \label{Omega}
\end{eqnarray}
where the subscript denotes differentiation with respect to the
radial coordinate. The condition $V_{eff,\;rr}(r)=0$ provides the
marginally stable orbit $r_{ms}$ (or the innermost stable circular
orbit), which can be determined for any explicit expression of the
function $\Phi(r)$.

\section{Electromagnetic signatures of accretion disks in
wormhole geometries}\label{sec:III}

Consider the static and spherically symmetric metric given by
\begin{equation}
ds^2=-e ^{2\Phi(r)}\,dt^2+\frac{dr^2}{1- b(r)/r}+r^2 \,(d\theta
^2+\sin ^2{\theta} \, d\phi ^2) \label{whmetric}\,,
\end{equation}
which describes a wormhole geometry with two identical,
asymptotically flat regions joined together at the throat $r_0>0$.
$\Phi(r)$ and $b(r)$ are arbitrary functions of the radial
coordinate $r$, denoted as the redshift function and the shape
function, respectively. The radial coordinate has a range that
increases from a minimum value at $r_0$, corresponding to the
wormhole throat, to $\infty$.

To avoid the presence of event horizons, $\Phi (r)$ is imposed to
be finite throughout the coordinate range. At the throat $r_0$,
one has $b(r_0)=r_0$, and a fundamental property is the
flaring-out condition given by $(b'r-b)/b^2<0$, which is provided
by the mathematics of embedding~\cite{Morris:1988cz}.

An interesting feature of wormhole geometries are their
repulsive/attractive character \cite{Lobo:2006ue}. To verify this,
consider the four-velocity of a static observer, at rest at
constant $r, \theta, \phi$, given by $u^{\mu}=dx^{\mu}/{d\tau}
=(u^{\,t},0,0,0)=(e^{-\Phi(r)},0,0,0)$. The observer's
four-acceleration is $a^{\mu} =u^{\mu}{}_{;\nu}\,u^{\nu}$, so that
taking into account metric (\ref{whmetric}), we have $a^t=0$ and
\begin{eqnarray}
a^r=\Gamma^{r}{}_{tt}\,\left(\frac{dt}{d\tau}\right)^2
    ={\Phi}_{,r}\,(1-b/r)\,.          \label{radial-acc}
\end{eqnarray}
Note that from the geodesic equation, a radially moving test
particle, which initially starts at rest, obeys the following
equation of motion
\begin{equation}
\frac{d^2r}{d\tau^2}=-\Gamma^{r}{}_{tt}
\left(\frac{dt}{d\tau}\right)^2=-a^r
\,.
      \label{radial-accel}
\end{equation}
$a^r$ is the radial component of proper acceleration that an
observer must maintain in order to remain at rest at constant
$r,\,\theta,\,\phi$. One may consider that the geometry is
attractive if $a^r>0$, i.e., observers must maintain an
outward-directed radial acceleration to keep from being pulled
into the wormhole; and repulsive if $a^r<0$, i.e., observers must
maintain an inward-directed radial acceleration to avoid being
pushed away from the wormhole. This distinction depends on the
sign of $\Phi_{,r}$, as is transparent from
Eq.~(\ref{radial-acc}). In particular, for a constant redshift
function, $\Phi_{,r}(r)=0$, static observers are also geodesic.
Thus, the convention used is that $\Phi_{,r}(r)$ is positive for
an inwardly gravitational attraction, and negative for an outward
gravitational repulsion.

The above analysis is of particular interest for particles moving
in circular orbits around wormholes, which are unstable for the
specific case of $\Phi_{,r}(r)<0$, due to the outward gravitational
repulsion. Therefore, we only consider the case of $\Phi_{,r}(r)>0$,
and in particular the example of
\begin{equation}
\Phi(r)=-\frac{r_0}{r}
\end{equation}
is considered throughout this work.

\begin{figure*}[t]
\centering
  \includegraphics[width=2.93in]{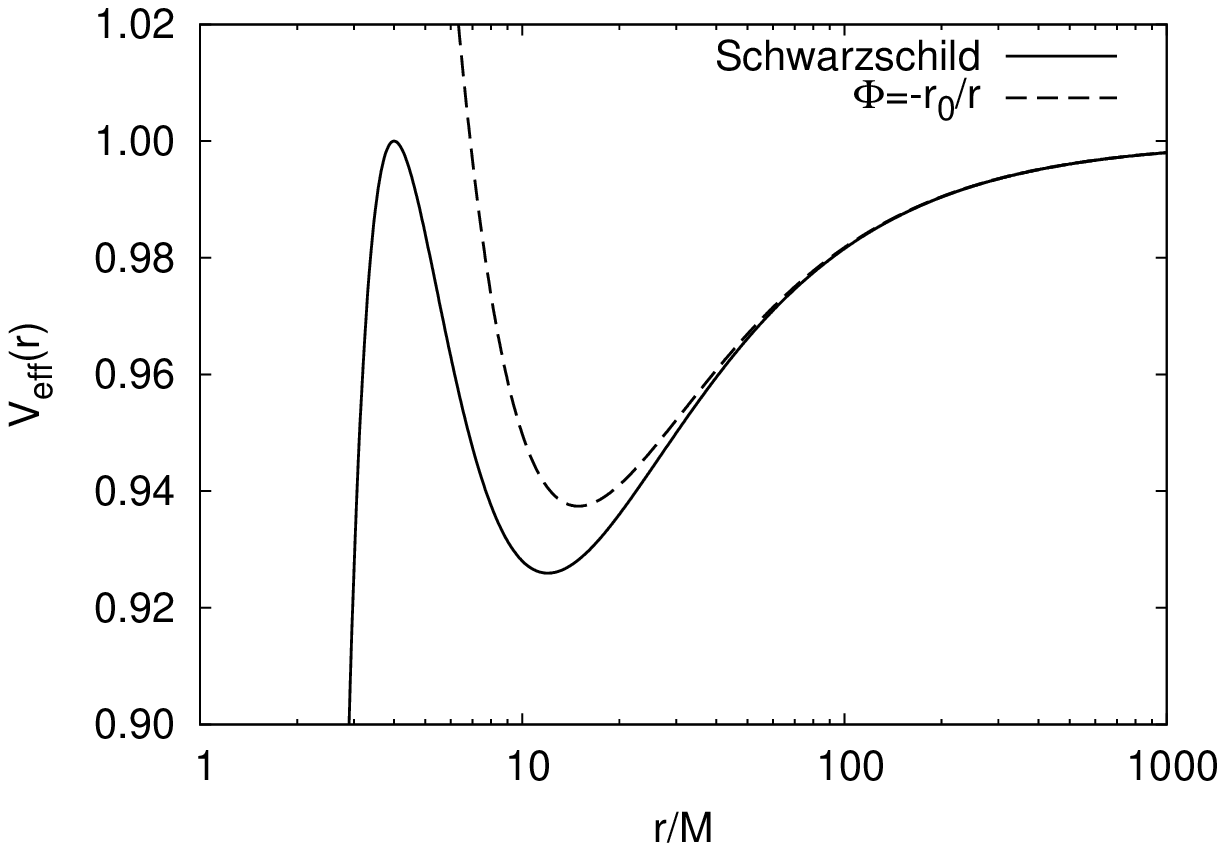}
  \includegraphics[width=2.93in]{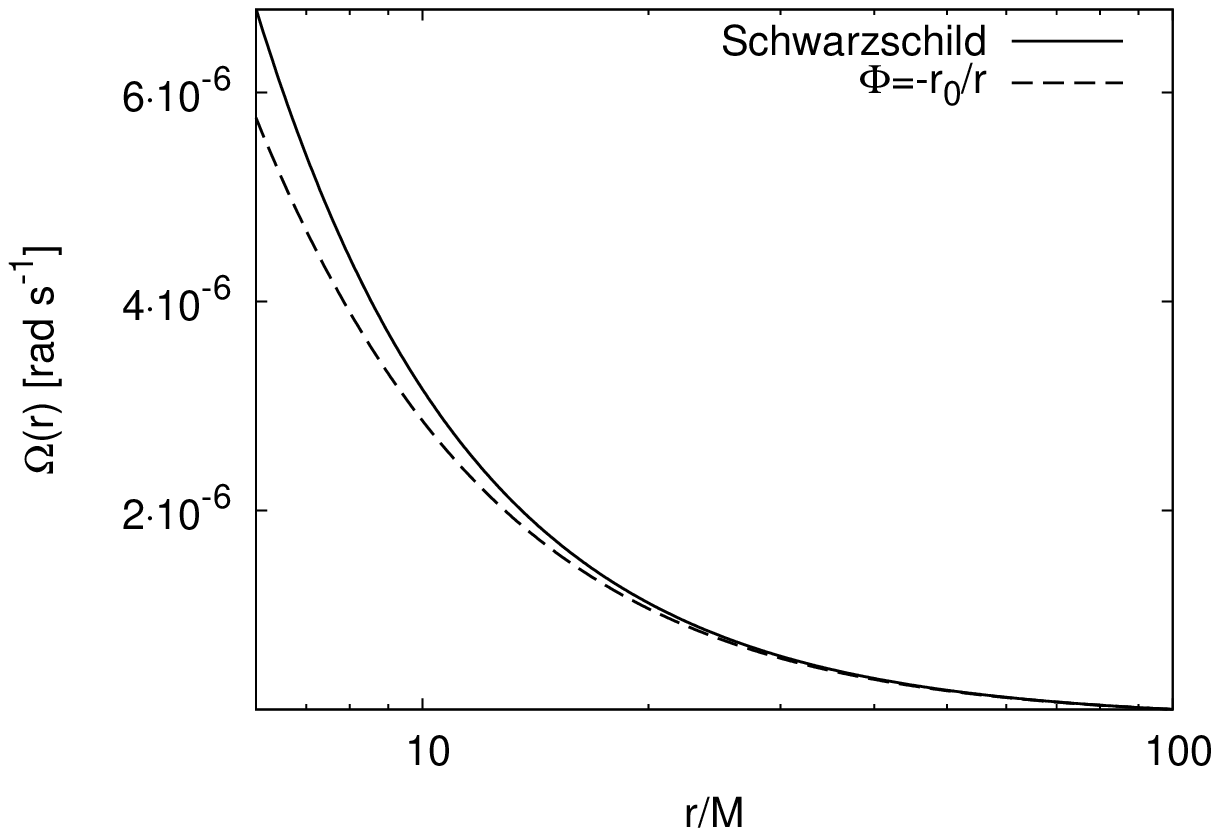}
  \includegraphics[width=2.93in]{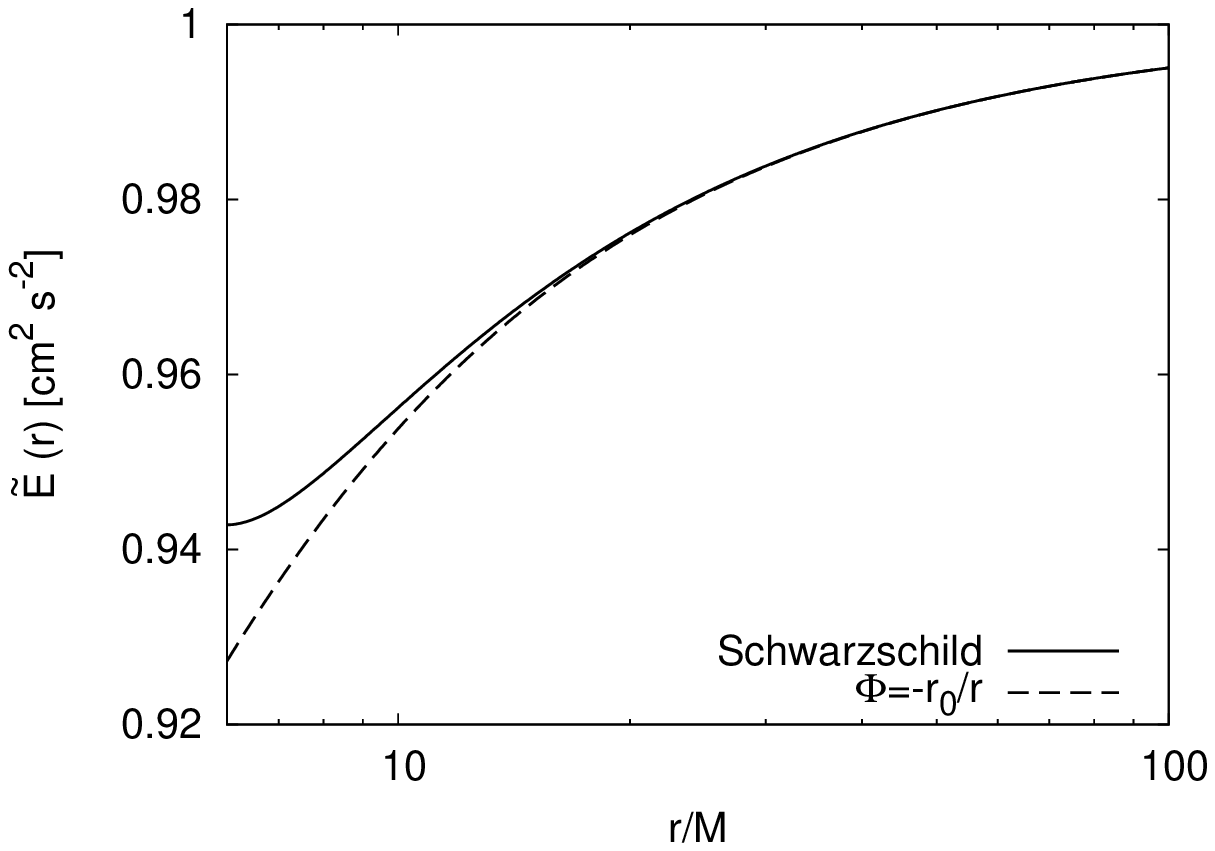}
  \includegraphics[width=2.93in]{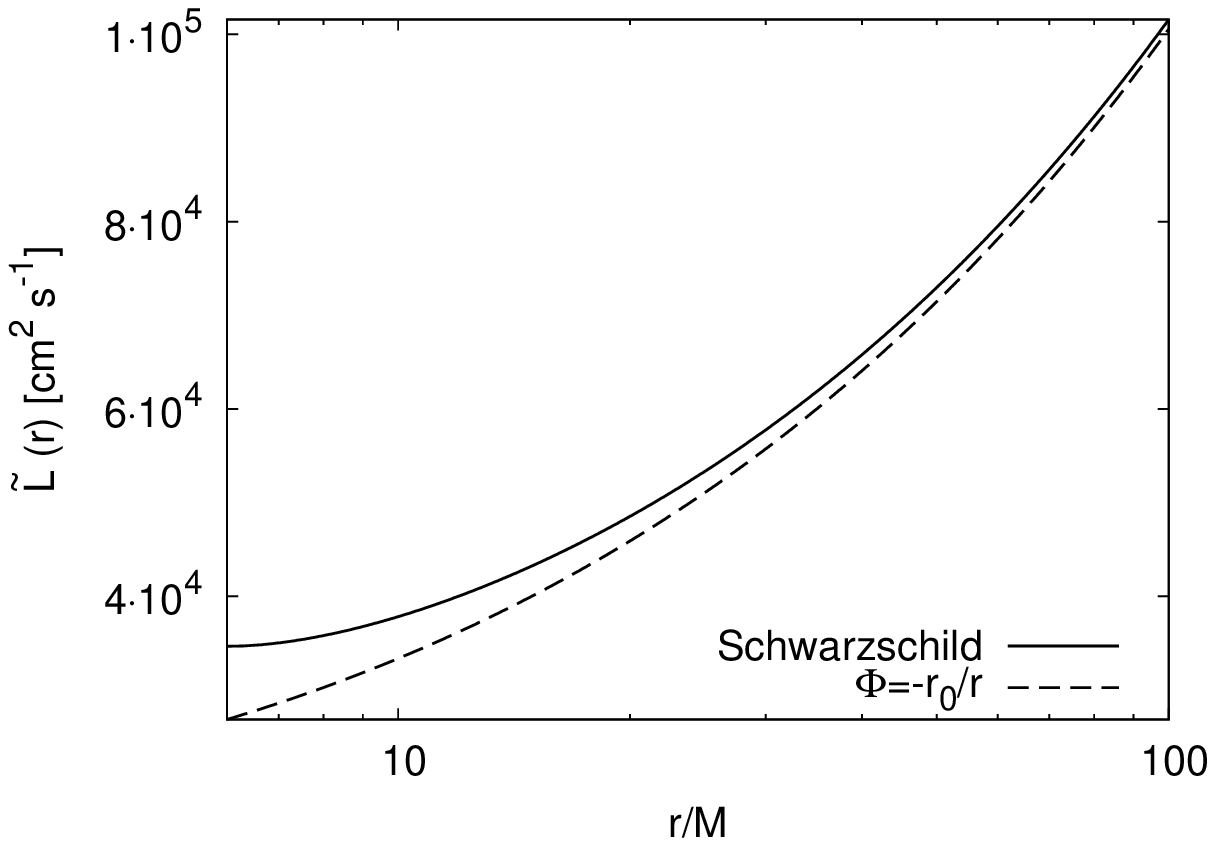}
  \caption{The effective potential $V_{eff}(r)$ (top left hand),
  the angular velocity $\Omega$ (top right hand), the specific
  energy $\widetilde{E}$ (bottom left hand) and the specific
  angular momentum $\widetilde {L}$ (bottom right hand) of the
  orbiting particles for a wormhole of a total mass
  $M=0.06776M_{\odot}$ and for the specific angular momentum
  $\widetilde{L}=4M$. The specific case of $\Phi=-r_0/r$ is
  considered, and is compared with the Schwarzschild solution,
  which is depicted as the solid line.}
  \label{Fig:whpotential}
\end{figure*}

The effective potential, given by Eq. (\ref{V2}), which determines
the geodesic motion of the test particles in the equatorial plane
of the metric given by Eq.~(\ref{whmetric}), is rewritten here for
convenience
\begin{equation}
V_{eff}(r)\equiv
e^{2\Phi}\left(1+\frac{\widetilde{L}}{r^2}^2\right)\;. \label{V2b}
\end{equation}
The specific case of $\Phi(r)=-r_0/r$ is depicted in
Fig.~\ref{Fig:whpotential}, for a wormhole with a total mass of
$M=0.06776M_{\odot}$ or $r_0=GM/c^2´=10^4$cm, and for particles
orbiting with the specific angular momentum of $\widetilde{L}=4M$.
The effective potential of the Schwarzschild geometry, depicted as
the solid line, is plotted for comparison.

In Fig. ~\ref{Fig:whpotential}, we also compare the angular
velocity $\Omega$, the specific energy $\widetilde{E}$ and the
specific angular momentum $\widetilde {L}$ of the orbiting
particles for the two types of the potential. Since these
quantities, given by Eqs. (\ref{E})-(\ref{Omega}), depend only on
the metric function $\Phi(r)$, they are identical for each
wormhole geometry, independently of the choice of the shape
function $b(r)$. The only quantity in the flux integral (\ref{F})
which has a dependence of $b(r)$ is the metric determinant. For
the invariant volume element of the various types of wormholes we
have
\begin{equation}
\sqrt{-g}=\frac{r^2e^{\Phi}}{\sqrt{1-b(r)/r}}\;,\label{sqrtmg}
\end{equation}
which can give different flux values in computing Eq. (\ref{F})
for different choices of $b(r)$. As seen in Fig.~\ref{Fig:whpotential},
for the wormhole geometry under study the
orbiting particles gain somewhat less angular velocity, specific
energy and specific angular momentum than in the Schwarzschild
potential.

\begin{figure*}
\centering
  \includegraphics[width=2.93in]{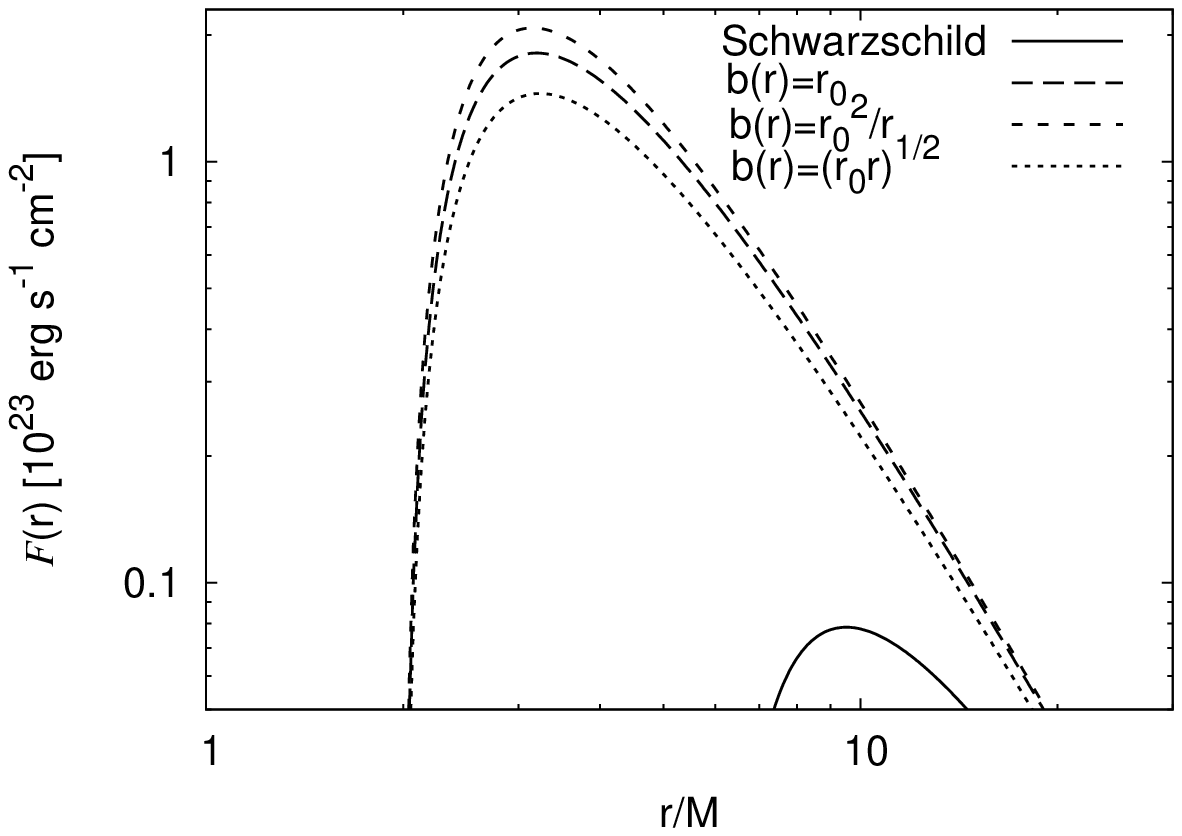}
\hspace{0.2in}
  \includegraphics[width=2.93in]{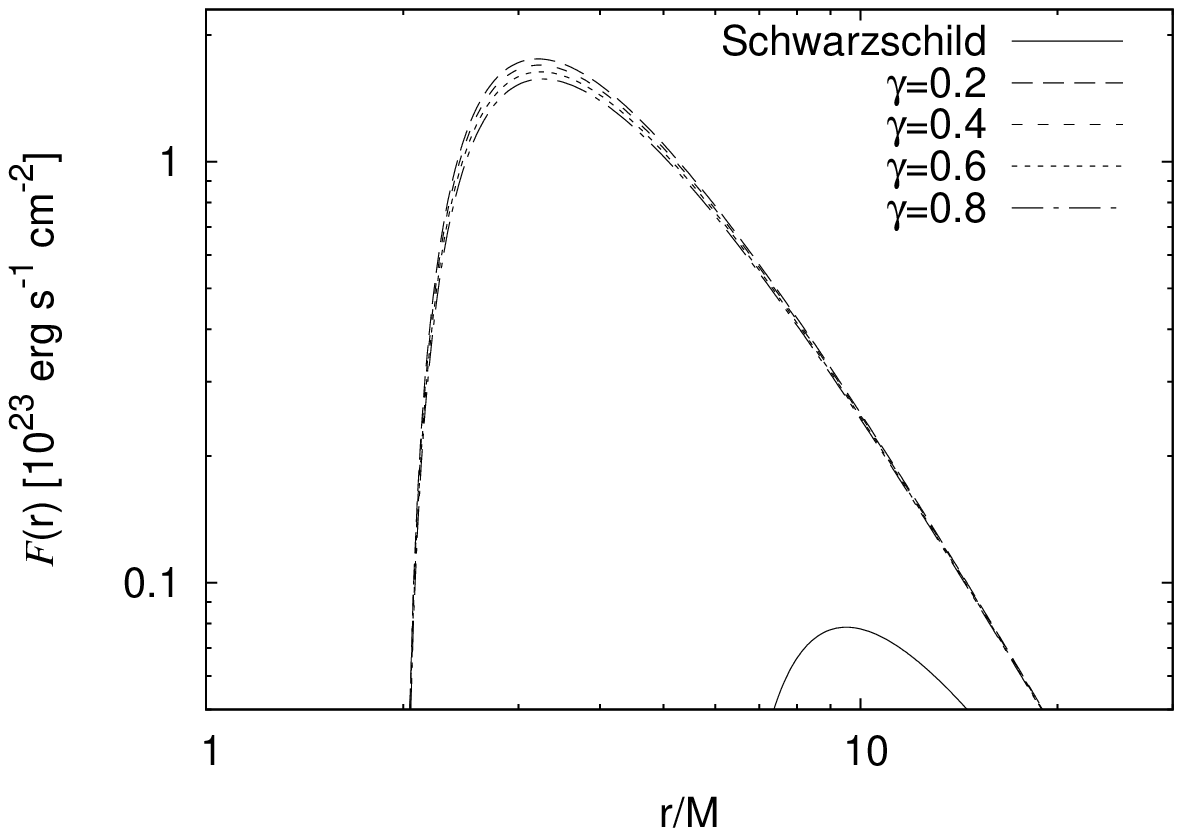}
\caption{The time averaged flux ${\cal F}(r)$ radiated by the disk
for a wormhole of an effective mass $M=0.06776M_{\odot}$ and a
mass accretion rate $\dot{M}_0=10^{-12}M_{\odot}/{\rm yr}$.
Different shape functions were evaluated, namely, in the left plot
the cases chosen were: $b(r)=r_0$ (the long dashed line),
$b(r)=r_0^2/r$ (the short dashed line) and $b(r)=(r_0r)^{1/2}$
(the dotted line), respectively. The flux emitted by a disk around
a Schwarzschild black hole is plotted with a solid line. The right
plot depicts the case of $b(r)=r_0+\gamma r_0(1-r_0/r)$, with
varying values of $\gamma$. }
  \label{Fig:whFlux}
\end{figure*}

\begin{figure*}
\centering
  \includegraphics[width=2.93in]{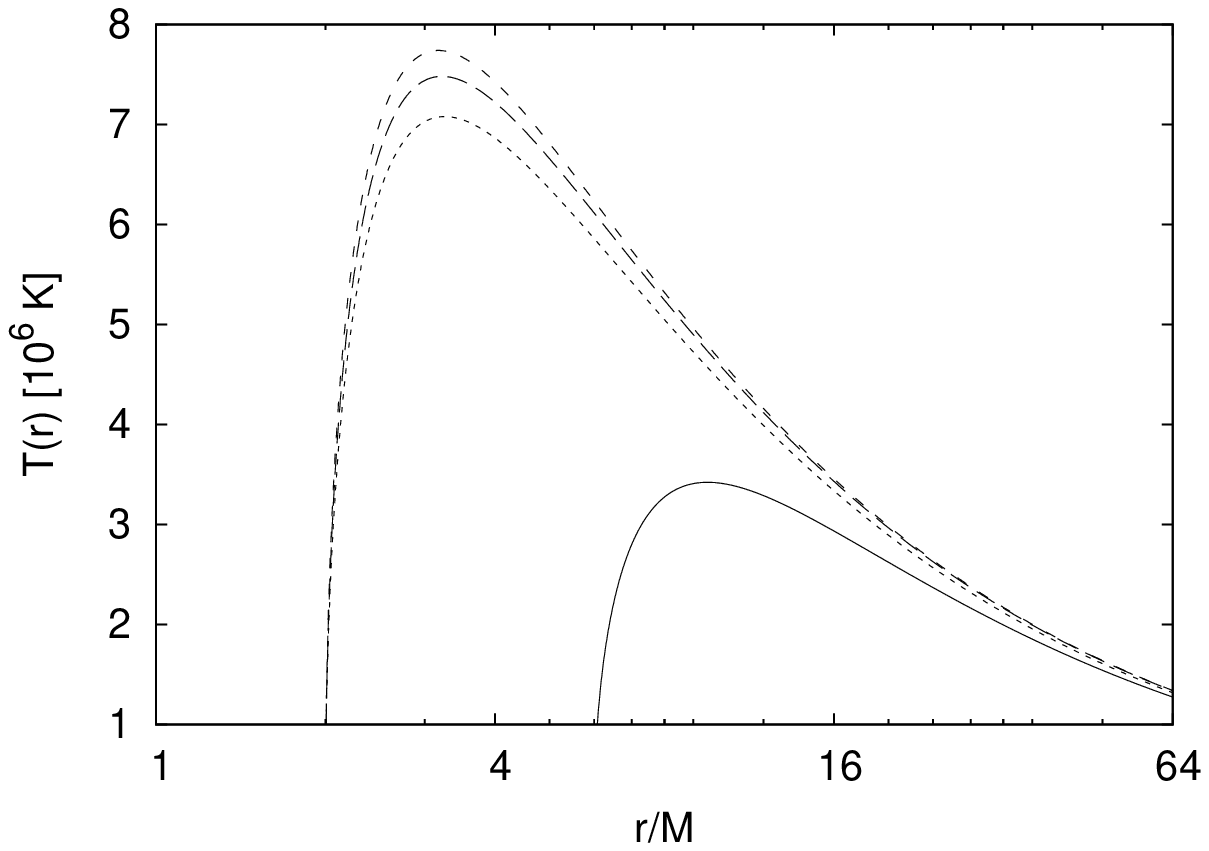}
\hspace{0.2in}
  \includegraphics[width=2.93in]{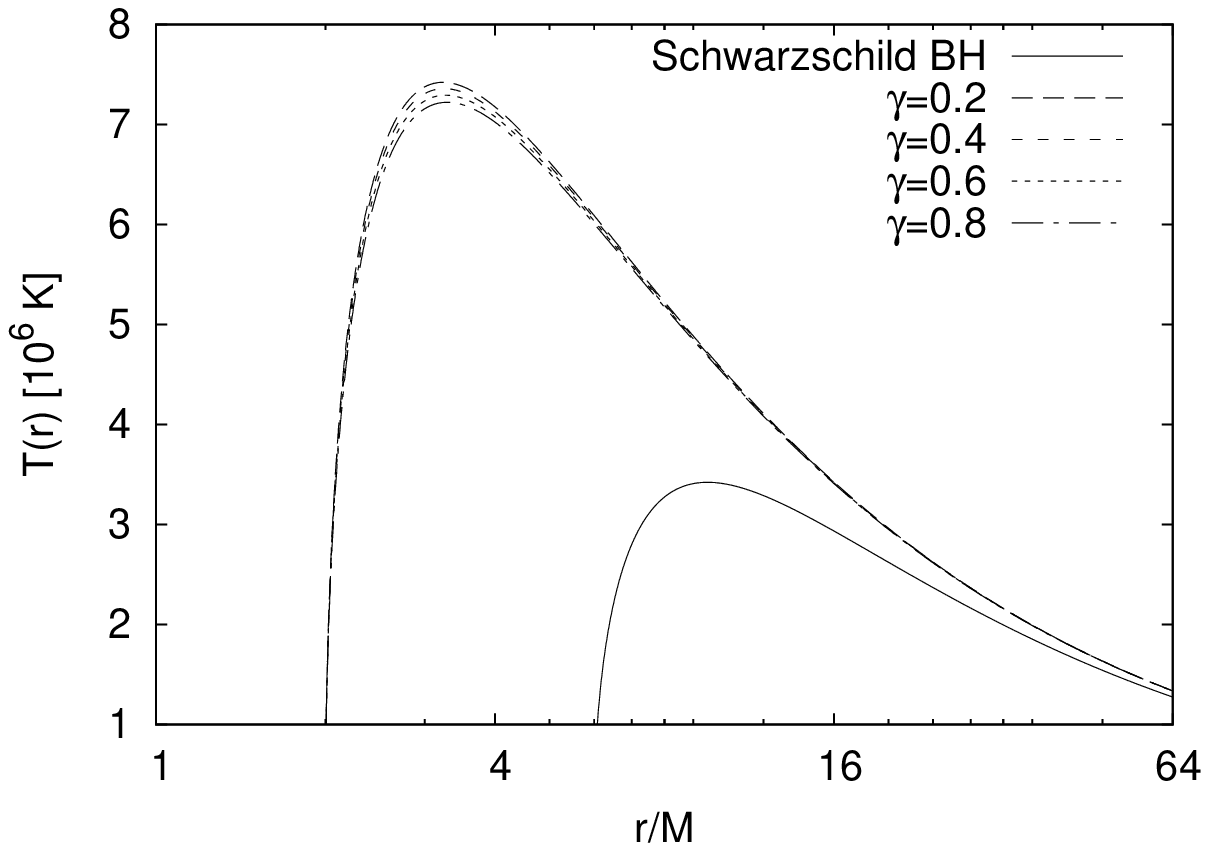}
\caption{Temperature distribution of the accretion disk for a
wormhole of an effective mass $M=0.06776M_{\odot}$ and a mass
accretion rate $\dot{M}_0=10^{-12}M_{\odot}/{\rm yr}$. Different
shape functions were evaluated, namely, in the left plot the
chosen cases were: $b(r)=r_0$ (the long dashed line),
$b(r)=r_0^2/r$ (the short dashed line) and $b(r)=(r_0r)^{1/2}$
(the dotted line), respectively. The disk temperature profile for
a Schwarzschild black hole is plotted with a solid line. The right
plot depicts the case of $b(r)=r_0+\gamma r_0(1-r_0/r)$, with
varying values of $\gamma$. }
  \label{Fig:whTemp}
\end{figure*}

\begin{figure*}
\centering
  \includegraphics[width=2.93in]{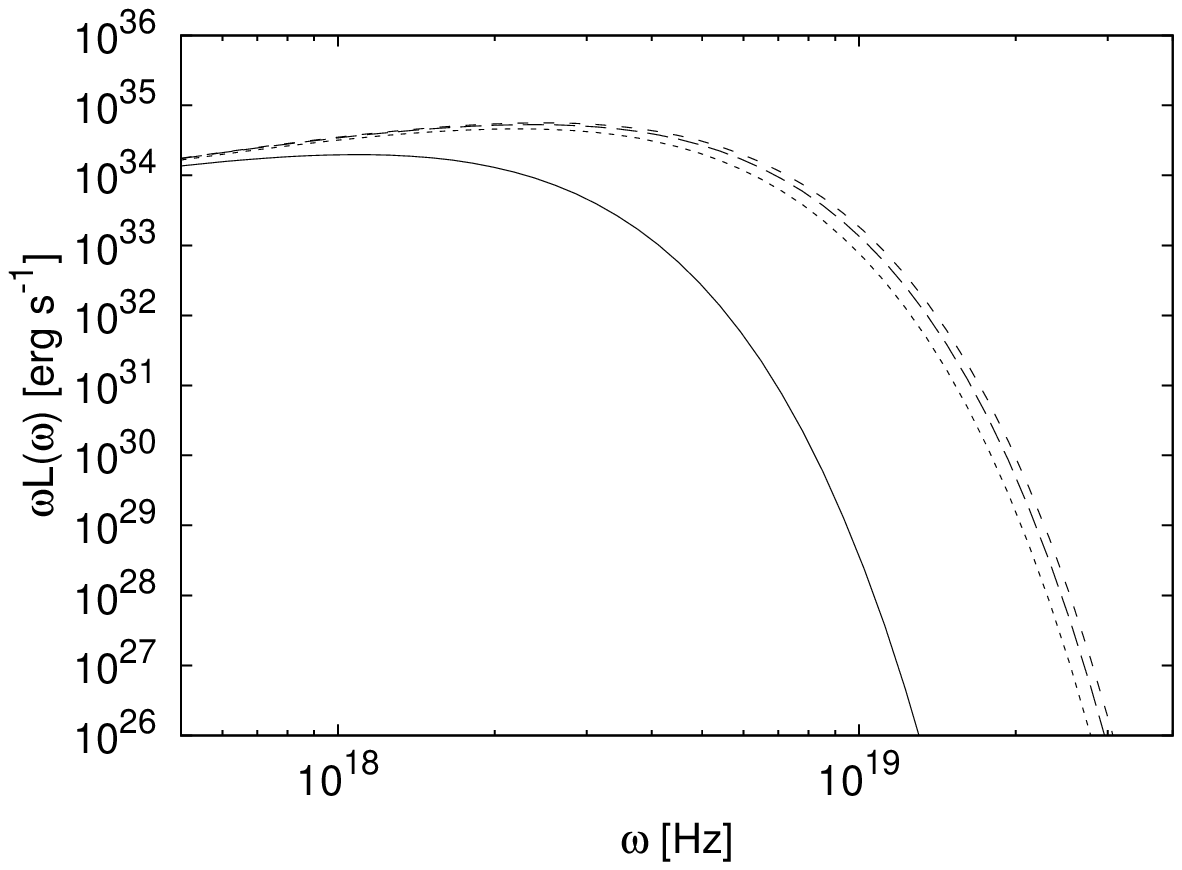}
\hspace{0.2in}
  \includegraphics[width=2.93in]{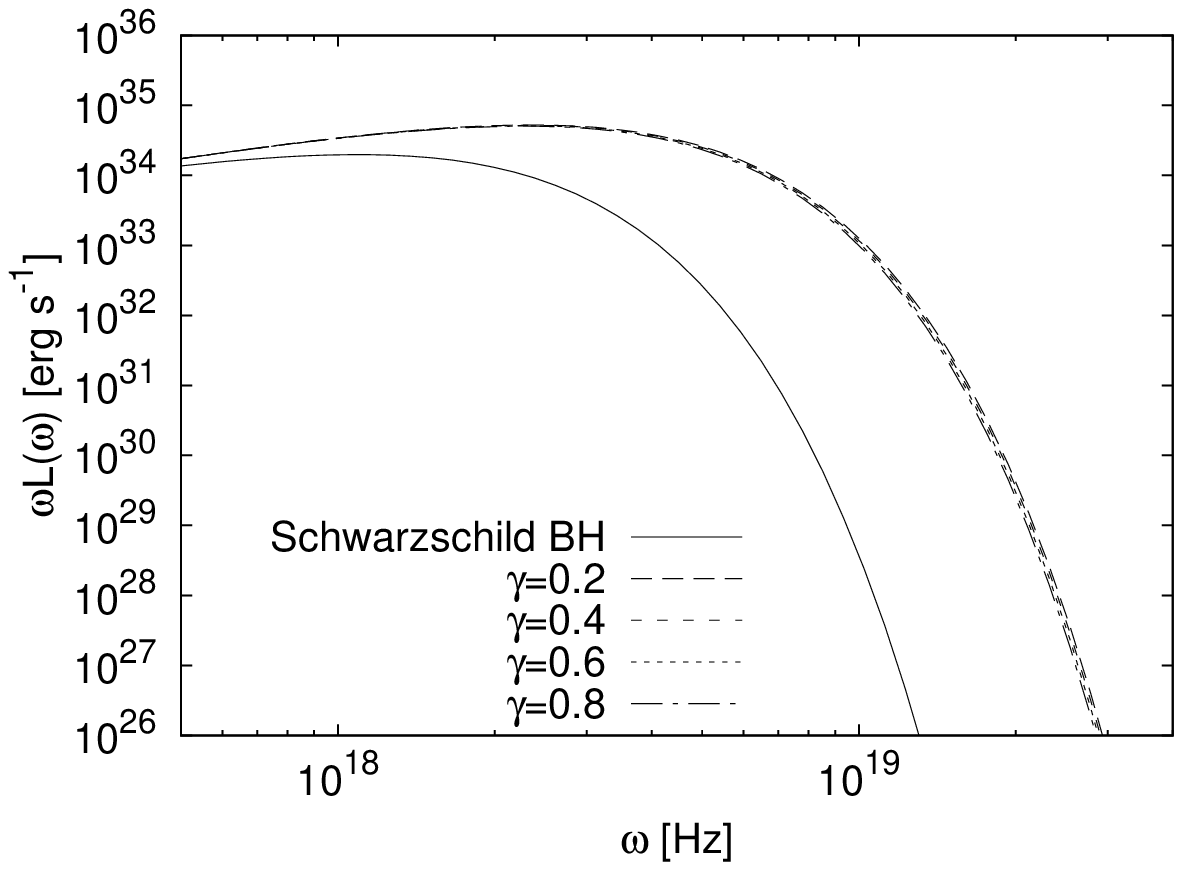}
\caption{The emission spectra $\omega L(\omega)$ of the accretion
disks for a wormhole of effective mass $M=0.06776M_{\odot}$ and a
mass accretion rate of $\dot{M}_0=10^{-12}M_{\odot}/{\rm yr}$.
Different shape functions were evaluated, namely, in the left plot
the chosen cases were: $b(r)=r_0$ (the long dashed line),
$b(r)=r_0^2/r$ (the short dashed line) and $b(r)=(r_0r)^{1/2}$
(the dotted line), respectively. The flux emitted by a disk around
a Schwarzschild black hole is plotted with a solid line. The right
plot depicts the case of $b(r)=r_0+\gamma r_0(1-r_0/r)$, with
varying values of $\gamma$.}
 \label{Fig:whspectra}
\end{figure*}

In Figs.~\ref{Fig:whFlux}--\ref{Fig:whspectra}, we plot the energy
flux, the disk temperature and the emission spectra $\omega
L(\omega)$ emitted by the accretion disk with a mass accretion
rate of $\dot{M}_0=10^{-12}M_{\odot}/{\rm yr}$ for various
wormhole geometries. In particular, in the left plots, of the
respective figures, we have used the following shape functions:
\begin{equation}
b(r)=r_0\,, \qquad  b(r)=\frac{r_0^2}{r}\,, \qquad
b(r)=\sqrt{rr_0}\,,
   \label{form1}
\end{equation}
and in the right plot, the shape function
\begin{equation}
b(r)=r_0+\gamma r_0\left(1-\frac{r_0}{r}\right)\,,
   \label{form2}
\end{equation}
has been used, with $0<\gamma<1$. Note that not only $\Omega$
$\widetilde{E}$ and $\widetilde{L}$ have smaller values for the
wormhole geometry under study than in the case of static black
holes but the invariant volume element (\ref{sqrtmg}) also takes
rather small values for any choice of the function $b(r)$.
Consequently more energy is radiated away than in the
Schwarzschild potential, which is reflected in
Fig.~\ref{Fig:whFlux}.

The  marginally stable orbit is located at the radius $2r_0$ for
the wormholes, whereas we have $r_{ms}=6r_0$ for the Schwarzschild
black hole. As a result, the inner edge of the disk is closer to
the throat of the wormhole than to the event horizon of the black
hole, comparing any wormhole and black hole with the same
geometrical mass $r_0$. The wormhole efficiency $\epsilon$ of the
conversion of the accreted mass into radiation is $0.1422$ which
is much higher than $\epsilon=0.0572$ for the black hole. Although
14\% is still less than the 40\% obtained for Kerr black holes, it
demonstrates that static wormholes provide a more efficient
accretion mechanism to convert the gravitational energy into disk
luminosity than the mass accretion driven by static black holes.

These effects in the disk radiation can also be observed in the
radial profiles of temperature corresponding to the flux
distributions, shown in Fig.~\ref{Fig:whTemp}, and in the emission
spectrum $\omega  L(\omega)$ of the accretion disks, which are
plotted in Fig.~\ref{Fig:whspectra}.

Thus, for the static and spherically symmetric solutions, we
conclude that the specific signatures, that appear in the
electromagnetic spectrum, lead to the possibility of
distinguishing wormhole geometries from the Schwarzschild solution
by using astrophysical observations of the emission spectra from
accretion disks.

\section{Discussions and final remarks}\label{sec:concl}

In the present paper, we have studied thin accretion disk models
applied to the study of static and spherically symmetric wormhole
geometries, and have carried out an analysis of the properties of
the radiation emerging from the surface of the disk. In classical
general relativity, wormholes are supported by exotic matter,
which involves a stress energy tensor that violates the null
energy condition. Thus, it has proved interesting to analyze the
properties, namely, the time averaged energy flux, the disk
temperature and the emission spectra of the accretion disks around
these wormholes comprising of exotic matter. For the static and
spherically symmetric wormhole geometries under study, we have
verified that the potential well is higher than the Schwarzschild
potential, and consequently more energy is radiated away. These
effects in the disk radiation were also observed in the radial
profiles of temperature corresponding to the flux distributions,
and in the emission spectrum $\omega L(\omega)$ of the accretion
disks. Thus, for these solutions, we conclude that the specific
signatures, that appear in the electromagnetic spectrum, lead to
the possibility of distinguishing wormhole geometries from the
Schwarzschild solution by using astrophysical observations of the
emission spectra from accretion disks.

In this context, observations in the near-infrared (NIR) or X-ray
bands have provided important information about the spin of the
black holes, or the absence of a surface in stellar type black
hole candidates. In the case of the source Sgr A$^*$, where the
putative thermal emission due to the small accretion rate peaks in
the near infrared, the results are particularly robust. However,
up to now, these results have confirmed the predictions of general
relativity mainly in a qualitative way, and the observational
precision achieved cannot distinguish between the different
classes of compact/exotic objects that appear in the theoretical
framework of general relativity \cite{YuNaRe04}. However,
important technological developments may allow to image black
holes and other compact objects directly \cite{BrNa06a}. A
background illuminated black hole will appear in a silhouette with
radius $\sqrt{27}GM/c^2$, with an angular size of roughly twice
that of the horizon, and may be directly observed. With an
expected resolution of $20$ $\mu $as, submillimeter very-long
baseline interferometry (VLBI) would be able to image the
silhouette cast upon the accretion flow of Sgr A$^*$, with an
angular size of $\sim 50$ $\mu $as, or M87, with an angular size
of $\sim 25$ $\mu $as. For a black hole embedded in an accretion
flow, the silhouette will generally be asymmetric regardless of
the spin of the black hole. Even in an optically thin accretion
flow an asymmetry will result from special relativistic effects
(aberration and Doppler shifting). In principle, detailed
measurements of the size and shape of the silhouette could yield
information about the mass and spin of the central black hole, and
provide invaluable information on the nature of the accretion
flows in low luminosity galactic nuclei.

We suggest that by using the same imaging technique, which gives
the physical/geometrical properties of the silhouette of the
compact object cast upon the accretion flows on the compact
objects, would be able to provide clear observational evidence for
the existence of wormholes, and differentiate them from other
types of compact general relativistic objects. We conclude by
pointing out that specific signatures appear in the
electromagnetic spectrum of the thin accretion disks around
wormholes, thus leading to the possibility of detecting and
distinguishing wormhole geometries by using astrophysical
observations of the emission spectra from accretion disks. It is
also interesting to generalize the analysis carried out in this
work to stationary and axially symmetric wormhole spacetimes, and
compare the emission spectra to the Kerr solution. Work along
these lines is currently underway.

\section*{Acknowledgments}

The work of TH is supported by an RGC grant of the government of
the Hong Kong SAR. ZK was supported by the Hungarian Scientific
Research Fund (OTKA) grant no.69036. FSNL was funded by
Funda\c{c}\~{a}o para a Ci\^{e}ncia e a Tecnologia (FCT)--Portugal
through the grant SFRH/BPD/26269/2006.

\end{document}